\documentclass {article}
\usepackage{moreverb,url}

\usepackage[colorlinks,bookmarksopen,bookmarksnumbered,citecolor=red,urlcolor=red]{hyperref}

\newcommand\BibTeX{{\rmfamily B\kern-.05em \textsc{i\kern-.025em b}\kern-.08em
T\kern-.1667em\lower.7ex\hbox{E}\kern-.125emX}}

\usepackage{siunitx}
\usepackage{amsmath} 
\usepackage{amssymb}
\usepackage{float}
\usepackage{booktabs}
\usepackage{array}
\usepackage{url}
\usepackage{graphicx}
\usepackage{epstopdf}
\usepackage{pdflscape}     % For landscape pages
\usepackage{float}         % For [H] float specifier
\usepackage{booktabs}      % For \toprule, \midrule, \bottomrule
\usepackage{graphicx}      % For \resizebox
\usepackage[most]{tcolorbox}
\usepackage{xcolor}
\usepackage{mathrsfs}
\usepackage{rotating}
\usepackage{relsize}
\usepackage{tabularx}
\usepackage{tikz}
\usetikzlibrary{decorations.pathreplacing, arrows.meta, positioning}

\usepackage[numbers]{natbib}
\usepackage{tfrupee}
\usepackage{authblk}
\newcommand{\keywords}[1]{\par\addvspace\baselineskip
\noindent\textbf{Keywords:}\enspace\ignorespaces#1}
\usepackage{amsthm}
\theoremstyle{remark} % Sets the style to use italic title and roman body font
\usepackage{amssymb}
\usepackage{ragged2e}
\usepackage{geometry}
\title{Order-Flow Filtration and Directional Association with Short-Horizon Returns}

\begin{document}

\author[1,2]{Aditya Nittur Anantha}
\author[1]{Shashi Jain}
\author[3]{Prithwish Maiti}

\affil[1]{Department of Management studies, Indian Institute of Science, Bengaluru, Karnataka, 560012, India}
\affil[2]{SigmaQuant Technologies Pvt. Ltd, Bengaluru, Karnataka, 560066, India}
\affil[3]{AlgoQuant Technologies Ltd., Bengaluru, Karnataka, 560066, India}            

\maketitle

\begin{abstract}
Electronic markets generate dense order flow with many transient orders, which degrade directional signals derived from the limit order book (LOB)~\cite{hasbrouck2013latency,dahlstrom2024cancellations}. We study whether simple structural filters on order lifetime, modification count, and modification timing sharpen the association between order book imbalance (OBI) and short-horizon returns in BankNifty index futures, where unfiltered OBI is already known to be a strong short-horizon directional indicator \cite{cont2014impact,cont2014price,hanke2015ofi}. The efficacy of each filter is evaluated using a three-step diagnostic ladder: contemporaneous correlations, linear association between discretised regimes, and Hawkes event-time excitation between OBI and return regimes. Our results indicate that filtration of the aggregate order flow produces only modest changes relative to the unfiltered benchmark. By contrast, when filters are applied on the parent orders of executed trades, the resulting OBI series exhibits systematically stronger directional association. Motivated by recent regulatory initiatives to curb noisy order flow, we treat the association between OBI and short-horizon returns as a policy-relevant diagnostic of market quality. We then compare unfiltered and filtered OBI series, using tick-by-tick data from the National Stock Exchange of India, to infer how structural filters on the order flow affect OBI-return dynamics in an emerging market setting.
\end{abstract}


\section{Introduction}

Modern electronic markets generate order flow characterised by rapid order updates~\cite{hasbrouck2013latency}. This dense order flow has resulted in increased adoption of automation and a significant reduction in latency~\cite{budish2015armsrace}. Microstructure studies show that most limit orders are cancelled without execution and that cancellation behaviour is closely related to execution risk and queue position~\cite{dahlstrom2024cancellations}. Recent work on fleeting orders and flickering quotes documents that a non-trivial fraction of visible depth disappears almost immediately after it appears~\cite{easley2016}. Large, short-lived orders can be used strategically to create spurious depth and mislead agents that condition on the visible limit order book~\cite{wang2017spoofing}. Taken together, the current literature suggests that ephemeral orders whose genuine trading intent is ambiguous, occupy a large proportion of the total order flow. 

In the presence of such fleeting orders, the association between the visible order flow and realised returns becomes harder to interpret empirically because a substantial fraction of messages does not lead to execution~\cite{wang2017spoofing}. To address these concerns, several exchanges and regulators have introduced regulations that explicitly link the number of orders submitted to the number of resulting trades, in the form of fees or penalties on high order-to-trade ratios~\cite{cboeeurope2025otrmanual,eurex2023otrvolfactor,sec2023edgxotr}. These measures increase the cost of submitting large numbers of short-lived orders. The empirical literature studies their impact on standard proxies for market quality such as spreads, depth and volatility~\cite{friederich2015otrliquidity,jorgensen2018throttling,capelleblancard2017curbing,lepone2013chix,khomyn2021algosgonewild}.

In Indian markets, order-to-trade ratio based measures were introduced by the regulator in 2012-13 to curb excessive order submissions relative to executions~\cite{sebi2012algotrading, sebi2013algoupdate,sebi2020otr}. Existing empirical studies of such interventions evaluate their impact on market quality using proxies such as spreads, depth and realised volatility~\cite{aggarwal2023otrfee}. In continuation with the efforts to maintain price-efficiency, regulators have recently introduced order-based surveillance measures that penalise patterns of frequent modification and cancellation without execution. In 2021-22, Indian equity derivatives markets~\cite{nse2021pnc,nse2022pncupdate} implemented the `Persistent Noise Creator' framework with penalties for relatively higher use of price-changing modifications and cancellations~\cite{nse2021pnc,nse2022pncupdate}. These penalties are computed by monitoring the order flow of each client on a per-contract basis. 

However, fleeting orders can arise for several reasons. During periods of volatile market activity, liquidity providers repeatedly adjust or cancel standing limit orders to manage adverse selection. The literature also documents episodes of manipulative intent in which large orders are placed and cancelled in quick succession to create an illusion of depth or directional interest. This leaves open the question of how to evaluate the informational content of the order flow that remains when these penalties are in place.

Among the liquidity estimation methods present in the rich literature on market microstructure, order flow or order book imbalance (OBI), the net pressure between buy and sell interest at or near the best quotes, is strongly associated with short-horizon price formation. Cont, Kukanov and Stoikov~\cite{cont2014impact,cont2014price} show that, over short intervals, price changes are mainly driven by order flow imbalance at the best bid and ask, with a robust linear relation that is stable across a large cross-section of U.S. stocks and across intraday sampling scales. Hanke and Weigerding~\cite{hanke2015ofi} document imbalance effects on German equities, and more recent work uses order-flow imbalance as an input to trading models that extract alpha at multiple horizons from the limit order book~\cite{kolm2023deepofi,yagi2023complexity}. In this literature, OBI has become a natural primitive for constructing short-horizon directional signals. Following this strand of investigation, we ask whether the association between OBI and realised short-horizon returns can be used as a diagnostic for market quality in markets where order-based penalties and surveillance are already in place.

We propose a three-step diagnostic framework that is suited to evaluating such regulations and illustrate it on BANKNIFTY futures data from the Indian National Stock Exchange (NSE) where order surveillance has recently been tightened. First, we examine the linear association between the OBI series and short-horizon returns using Pearson correlations at multiple lags. Second, we discretise both imbalance and returns into regimes and ask how the empirical distribution of future return regimes depends on the current imbalance regime, treating regimes as states of a counting-process representation. Third, we promote regime transitions to point-process events and fit multivariate Hawkes models to the joint dynamics of imbalance and return regimes. In this last step of the diagnostic ladder, we interpret fitted Hawkes kernel norms and branching ratios as diagnostics of how strongly past imbalance events excite future return events, in line with the use of Hawkes processes to study endogeneity and reflexivity in financial markets~\cite{bowsher2007modelling,bacry2015hawkes,hardiman2013reflexivity,filimonov2015hawkes,hawkes1971spectra}.

In this setting, our contribution is deliberately diagnostic. Using detailed order-level data for a liquid BANKNIFTY index futures contract (see Table~\ref{tab:event_level_filtration}), we implement the three-step ladder described above on both unfiltered and filtered imbalance signals. We then compare their behaviour across three trading days, selected to span the early, middle and late parts of a monthly futures expiry cycle and to contrast relative levels of market activity. We define three simple, lifecycle-based filters at the order level (order lifetime, modification count and time between successive modifications). We apply the filters directly to the standing LOB and indirectly via the parent orders of executed trades, and recompute order-based and trade-based imbalance series in each case. We then evaluate association with short-horizon returns through Pearson correlations, regime-based association and event-time excitation between OBI and return regimes encoded in the kernel norms of a parametrically specified Hawkes process.

Using this diagnostic ladder on the tick-by-tick data for BANKNIFTY index futures, we find that order flow filtration schemes do not consistently improve the alignment between order-based imbalance and future returns across days and horizons, despite the intuition from the cancellation and fleeting-order literature. However, when the same filters are applied via the parent orders of realised trades, the resulting filtered trade-based imbalance exhibits systematically stronger Hawkes cross-excitation kernel norms from imbalance to return regimes than its unfiltered counterpart. Within the limits of our sample and without specifying a forecasting or execution strategy, our results suggest that not all trades contribute equally to price formation. 

\subsection{Related literature}
\label{subsec:related_literature}
Our work relates to three strands of the market microstructure literature. The first concerns order flow imbalance and short-horizon price impact. Studies such as \cite{cont2014impact,cont2014price,huang2015qmf} show that relatively simple, unfiltered measures of net order flow already have substantial explanatory power for intraday returns. In this perspective, OBI acts as a compact summary of directional pressure, and much of the empirical focus is on how OBI is associated with prices across horizons and assets.

The second strand focuses on low-latency message traffic and the prevalence of fleeting orders. Evidence on cancellations, rapid order revisions, and quote flickering \cite{hasbrouck2013latency,easley2016,dahlstrom2024cancellations} suggests that a large share of the observed event stream reflects activity that does not culminate in execution. 

A third strand models event-time dynamics and reflexivity using point processes. Multivariate Hawkes specifications have been used to capture clustered excitation in order arrivals, trades, and price changes \cite{bowsher2007modelling,bacry2015hawkes}. Hawkes processes have also been used to study endogeneity and apparent criticality in high-frequency markets \cite{hardiman2013reflexivity,filimonov2015hawkes}. Related work on state-dependent intensities and regime structure \cite{morariu2022statedependent,li2023regimeswitching} underscores the role of latent states in shaping price formation from order flow interactions. 

In this paper we draw on these ideas in a restricted way: Hawkes processes are used as diagnostic tools to quantify excitation from imbalance regimes to return regimes under different filtration schemes, rather than as forecasting engines or fully specified structural models. We use the empirical findings on OBI as a diagnostic lens to analyse price formation in high-frequency pre- and post-filtration.

\section{Problem Description}
\label{sec:problem_description}
We begin by introducing the counting processes and derived statistics used to quantify directional order flow within fixed-length evaluation windows. Each quantity is defined using primitives derived from the limit order book.

\begin{tcolorbox}[colframe=black!60, colback=white!97!gray, boxrule=0.4pt, arc=2pt, left=4pt, right=4pt, top=4pt, bottom=4pt, enhanced, sharp corners=south]
\textbf{Definition: Cumulative Event Count.}

\smallskip
\noindent
Let \( N_t \) denote the cumulative count of all LOB events observed up to time \( t \):
\begin{align}
N_t := \sum_{t_i} \mathbb{I}\{t_i \leq t\} \label{eq:cumulative_count}
\end{align}
where each \( t_i \) is an event timestamp and \( \mathbb{I}\{\cdot\} \) is the indicator function.
\end{tcolorbox}

We next define the total number of events arriving in a given lookback window. This interval-level statistic forms the base upon which directional and filtered counts are built.

\begin{tcolorbox}[colframe=black!60, colback=white!97!gray, boxrule=0.4pt, arc=2pt, left=4pt, right=4pt, top=4pt, bottom=4pt, enhanced, sharp corners=south]
\textbf{Definition: Event Count over a Lookback Window.}

\smallskip
\noindent
For fixed horizon \( h \), the number of LOB events in the interval \( (\tau - h, \tau] \) is given by:
\begin{align}
\Delta N_{(\tau - h, \tau]} := \sum_{t_i} \mathbb{I}\{t_i \in (\tau - h, \tau]\} \label{eq:total_event_count}
\end{align}
\end{tcolorbox}

Each event in the sequence carries a directional label \( Y_i \in \{b, s\} \) identifying the active side. The next definition isolates directional contributions attributed to an event count.

\begin{tcolorbox}[colframe=black!60, colback=white!97!gray, boxrule=0.4pt, arc=2pt, left=4pt, right=4pt, top=4pt, bottom=4pt, enhanced, sharp corners=south]
\textbf{Definition: Directional Event Count.}

\smallskip
\noindent
The count of events on side \( Y \in \{b, s\} \) in the interval \( (\tau - h, \tau] \) is:
\begin{align}
\Delta N^Y_{(\tau - h, \tau]} := \sum_{t_i \in (\tau - h, \tau]} \mathbb{I}\{Y_i = Y\} \label{eq:directional_count}
\end{align}
All event types are included without filtering or weighting.
\end{tcolorbox}

Using these directional statistics, we define a normalized imbalance measure that reflects net order flow pressure within each window.

\begin{tcolorbox}[colframe=black!60, colback=white!97!gray, boxrule=0.4pt, arc=2pt, left=4pt, right=4pt, top=4pt, bottom=4pt, enhanced, sharp corners=south]
\textbf{Definition: Order Book Imbalance.}

\smallskip
\noindent
The order book imbalance (OBI) over the window \( (\tau - h, \tau] \) is defined as:
\begin{align}
OBI(\tau, h) := \frac{ \Delta N^s_{(\tau - h, \tau]} - \Delta N^b_{(\tau - h, \tau]} }{ \Delta N^s_{(\tau - h, \tau]} + \Delta N^b_{(\tau - h, \tau]} } \label{eq:ofi}
\end{align}
where the denominator equals the \emph{total number of orders} arriving in the interval. When the denominator is zero, indicating no directional activity, the interval is excluded from analysis.
\end{tcolorbox}

To evaluate market response over the same interval, we construct a realized return from executed trades. This requires isolating the relevant subset of the event stream and transforming it into price points.
\begin{tcolorbox}[colframe=black!60, colback=white!97!gray, boxrule=0.4pt, arc=2pt, left=4pt, right=4pt, top=4pt, bottom=4pt, enhanced, sharp corners=south]
\textbf{Definition: Realized Return from Trade event \\ Filtration.}
\\
\noindent
Let the LOB event type space be defined as:
\begin{align}
\mathcal{E} := \{ \eta, T, M, C \} \label{eq:event_types}
\end{align}
where \( \eta \) denotes new order submission, \( T \) trade, \( M \) modification, and \( C \) cancellation.

Let \( \epsilon_t \in \mathcal{E} \) denote the event stream from $(\tau - h, \tau]$. Define a filtration that selects trade events within a lookback window:
\begin{align}
\mathcal{F}^{T}_{(\tau - h, \tau]} := \left\{ \epsilon_i \in \epsilon_t : t_i \in (\tau - h, \tau],\; \epsilon_i = T \right\} \label{eq:trade_filtration}
\end{align}

Define a transformation that extracts the traded prices at the first and last trade timestamps in $(\tau-h, \tau]$:
\begin{align}
\mathcal{G}(\mathcal{F}^{T}_{(\tau - h, \tau]}) := \left( p^{T}_{\underline{t}},\; p^{T}_{\overline{t}} \right) \label{eq:price_mapping}
\end{align}
where 
\begin{align}
\underline{t} \;:=\; \min \{\, t_i : \epsilon_i \in \mathcal{F}^T_{(\tau-h,\tau]} \,\}, 
\end{align} 
\begin{align}
\overline{t} \;:=\; \max \{\, t_i : \epsilon_i \in \mathcal{F}^T_{(\tau-h,\tau]} \,\}.
\end{align}

and \( p^{T}_{\underline{t}} \) and \( p^{T}_{\overline{t}} \) are the traded prices at the earliest and latest timestamps in \( \mathcal{F}^{T}_{(\tau - h, \tau]} \), respectively.

Then the realized return over the interval is defined as:
\begin{align}
\tilde{r}_{(\tau - h, \tau]} := \frac{p^{T}_{\overline{t}} - p^{T}_{\underline{t}}}{p^{T}_{\underline{t}}} \label{eq:realized_return}
\end{align}
If \( \mathcal{F}^{T}_{(\tau - h, \tau]} = \emptyset \), the return is undefined and the window is excluded from analysis.
\end{tcolorbox}

\begin{figure}[H]
\centering
\begin{tikzpicture}[xscale=13, yscale=13]

  % Time axis
  \draw[->, thick] (0.0, -0.25) -- (0.9, -0.25) node[below=8pt, anchor=east] {Time};
  \draw[dashed] (0.15, -0.40) -- (0.15, 0.1);
  \draw[dashed] (0.815, -0.40) -- (0.81, 0.1);
  \node[below] at (0.19, -0.25) {\( \tau - h \)};
  \node[below] at (0.8, -0.25) {\( \tau \)};
  \node[above=5pt] at (0.5, -0.3075) {\footnotesize Evaluation window \( (\tau - h, \tau] \)};

  % Buy events (lower dots) -- centers of former bars
  % Original rectangles: (0.25,0)-(0.27,0.4), (0.35,0)-(0.37,0.4), (0.50,0)-(0.52,0.4)
  \filldraw[blue!60] (0.26, -0.10) circle (0.3pt) node[below=2pt, black] {};
  \filldraw[blue!60] (0.36, -0.10) circle (0.3pt) node[below=2pt, black] {};
  \filldraw[blue!60] (0.51, -0.10) circle (0.3pt) node[below=2pt, black] {};
  \node[left, blue!60!black] at (0.24, -0.10) {\footnotesize Buys};
  % Event-type annotation for buys
  \node[above=2pt, blue!60!black] at (0.26, -0.10) {\tiny Trade};
  \node[above=2pt, blue!60!black] at (0.36, -0.10) {\tiny New Order};
  \node[above=2pt, blue!60!black] at (0.51, -0.10) {\tiny Modification};

  % Sell events (upper dots) -- centers of former bars
  % Original rectangles: (0.30,0.65)-(0.32,0.95), (0.60,0.65)-(0.62,0.95), (0.75,0.65)-(0.77,0.95)
  \filldraw[red!70] (0.31, 0.0) circle (0.3pt) node[above=2pt, black] {};
  \filldraw[red!70] (0.61, 0.0) circle (0.3pt) node[below=2pt, black] {};
  \filldraw[red!70] (0.76, 0.0) circle (0.3pt) node[below=2pt, black] {};
  \node[left, red!60!black] at (0.29, 0.0) {\footnotesize Sells};
  % Event-type annotation for sells
  \node[below=2pt, red!60!black] at (0.31, 0.0) {\tiny New Order};
  \node[below=2pt, red!60!black] at (0.61, 0.0) {\tiny Modification};
  \node[below=2pt, red!60!black] at (0.76, 0.0) {\tiny Trade };

  % Braces for directional counts (unchanged)
  \draw[decorate,decoration={brace,mirror}, thick] (0.24, -0.13) -- (0.53, -0.13)
    node[midway, below=4pt] {\footnotesize \( \Delta N^b_{(\tau - h, \tau]} \)};
  \draw[decorate,decoration={brace}, thick] (0.29, 0.03) -- (0.78, 0.03)
    node[midway, above=4pt] {\footnotesize \( \Delta N^s_{(\tau - h, \tau]} \)};

  % Realized return arrow and trade endpoints (annotate as T)
  \draw[->, thick] (0.26, -0.35) -- (0.76, -0.35);
  \node[below] at (0.5, -0.35) {\( \tilde{r}_{(\tau - h, \tau]} \)};
  \draw[fill=black] (0.26, -0.35) circle (0.3pt) node[below left] {\tiny \( p^{T}_{\underline{t}} \)};
  \node[above left=-1pt] at (0.26, -0.35) {\tiny \( T \)};
  \draw[fill=black] (0.76, -0.35) circle (0.3pt) node[below right] {\tiny \( p^{T}_{\overline{t}} \)};
  \node[above right=-1pt] at (0.76, -0.35) {\tiny \( T \)};

  % Optional compact legend (event types)
  % \node[draw, very thin, fill=white, align=left, anchor=north east] at (1.02,1.15) {\tiny
  %   \( \eta \): new order (arrival)\\
  %   \( T \): trade (used for \( \tilde{r} \))
  % };

\end{tikzpicture}

\caption{\footnotesize Visual depiction of the evaluation window \( (\tau - h, \tau] \) used for computing both order book imbalance and realized return. The OBI is based on the net directional event counts, while the return is computed using the first and last trade prices within the same window.}
\label{fig:obi_return_window}
\end{figure}

Figure~\ref{fig:obi_return_window} shows how we compute realised returns. In every window $(\tau - h, \tau]$, we follow the methodology proposed in \cite{nittur2025tradeobi} and sign all trade events as `Buy' or `Sell'. We then look for the first and last trade events, irrespective of the classified sign, and compute the net difference in traded prices for that window. The sign of the trade events is used subsequently to compute the OBI, when using signed trades as the LOB primitive.

In this study, we evaluate whether the statistical association between order book imbalance and contemporaneous price change, both computed over the same backward-looking interval, is strengthened after filtering the LOB event stream. Specifically, we assess whether filtering improves explanatory power in terms of (i) empirical correlation between \( OBI(\tau, h) \) and \( \tilde{r}_{(\tau - h, \tau]} \), and (ii) the structure of fitted excitation kernels in a parametric Hawkes process.

To this end, we introduce a class of filtration schemes that operate on the raw event sequence before computing any statistics. These schemes remove ephemeral or structurally unreliable events based on short lifetimes, rapid modifications, or flickering placement behavior near the top of the LOB. The resulting filtered event streams yield corresponding statistics, denoted \( OBI^{(f)}(\tau, h) \), which are used for comparative evaluation.

\begin{tcolorbox}[colframe=black!60, colback=white!97!gray, boxrule=0.4pt, arc=2pt, left=4pt, right=4pt, top=4pt, bottom=4pt, enhanced, sharp corners=south]
\textbf{Problem Statement: Evaluation of Filtration-Driven Signal Strength.}

\smallskip
\noindent
Let $\epsilon_t \in \mathcal{E}$ be the observed event stream. Let $\mathscr{F}^{(\cdot)}$ denote a filtration scheme applied to $\epsilon_t$, with superscript $(\cdot)$ indexing a filtering criterion.

Let $\tilde{r}_{(\tau - h, \tau]}$ be the realized return computed over each window as defined previously. Define a scoring functional:
\begin{align}
\mathcal{S}: \mathscr{F}^{(\cdot)} \times \tilde{r} \longmapsto \mathbb{R} \label{eq:score_def}
\end{align}
which quantifies the strength of association between filtered order book imbalance and realized return.

We consider a set of filtration schemes $\{ \mathscr{F}^{(1)}, \ldots, \mathscr{F}^{(k)} \}$, each yielding filtered statistics $OBI^{(i)}$ and a common return sequence $\tilde{r}$. Our objective is to determine whether:
\begin{align}
\mathcal{S}(\mathscr{F}^{(i)}, \tilde{r}) > \mathcal{S}(\mathscr{F}^{(0)}, \tilde{r}) \quad \text{for some } i \in \{1, \dots, k\} \label{eq:score_comparison}
\end{align}
where $\mathscr{F}^{(0)}$ denotes the unfiltered case.
\end{tcolorbox}

\section{Methodology}
\label{sec:methodology}

The methodology uses three components. First, we define structural quantities and filtration schemes \( \mathscr{F}^{(\cdot)} \) that remove short-lived or heavily revised orders from the raw tick stream \( \epsilon_t \). Second, we construct imbalance indicators from filtered and unfiltered data, including a trade-based benchmark. Third, we apply a three-layer diagnostic ladder that moves from value-level correlation to regime-based association, and then to Hawkes excitation norms, in order to quantify directional alignment between imbalance and realized returns.

Using filtered events, we reconstruct the LOB, compute directional features in the evaluation windows, and compare them to a reference LOB constructed from the unfiltered stream. The scoring functionals \( \mathcal{S}(\mathscr{F}, \tilde{r}) \) are then applied to quantify the association between directional features and returns across filtration schemes. OBI is used to identify directional regimes and interpret score behavior; however, it is not used directly as an input to the score. For notational clarity, a summary of all symbols is provided in Appendix Table~\ref{tab:global_notation_clean}.

\subsection{Structure and Basic Quantities}
We first introduce the core temporal and structural quantities defined at the order level. Let \( t_j^{(1)} \) denote the entry time of order \( j \), defined as the timestamp at which the order is first submitted to the LOB. Let \( t_j^{(2)} \) denote its exit time, defined as the instant at which the order leaves the LOB, either because it is cancelled or because it is fully executed. The total time in the LOB for order \( j \) is given by
\begin{align}
\mathcal{T}_j := t_j^{(2)} - t_j^{(1)}.
\end{align}
This quantity is used to identify ephemeral orders.

\begin{tcolorbox}[colframe=black!40!white, colback=gray!3, boxrule=0.3pt]
\textbf{Modeling Remark.}
We couple cancellations with trades to compute exit-time because, in our tick data and also in the literature~\cite{dahlstrom2024cancellations}, cancellations vastly outnumber executions. Adding the number of trades to the number of cancellations leaves the aggregate essentially unchanged. Additionally, both cancellations and trades affect the same terminal outcome: the removal of an order from the LOB. However, modifications do not. Modifications update the order in place and are therefore treated separately via \(M_j\) and \(\mathcal{M}_j\).
\end{tcolorbox}

In addition to the survival time defined above, we define two modification-based measures. Let \( M_j \) denote the number of modifications recorded for the order \( j \) during its lifetime. We define \( \mathcal{M}_j \) as the time between the last two modifications of order \( j \), provided it was modified at least twice. The quantity \( \mathcal{M}_j \) captures timing intensity in the order update stream and serves to flag aggressively re-priced orders.

These quantities form the basis for the filtration schemes presented next.

\subsection{Filtration Schemes}
\label{subsec:filters}

Let \( \epsilon_t \in \mathcal{E} \) denote the observed LOB event stream. We denote a filter specification using \( \mathscr{F}^{(\cdot)} \). We define three filters defined using the total survival lifetime, modification count, and duration between successive modifications.

\begin{tcolorbox}[colframe=black!60, colback=white!97!gray, boxrule=0.4pt, arc=2pt, left=4pt, right=4pt, top=4pt, bottom=4pt, enhanced, sharp corners=south]
\textbf{Definition: Order Lifetime Filter, $\mathscr{F}^{\mathcal{T}}(\epsilon_t)$}

\smallskip
\noindent
Let \( t_j^{(1)} \) and \( t_j^{(2)} \) denote the entry and exit times of order \( j \), and define \( \mathcal{T}_j := t_j^{(2)} - t_j^{(1)} \) as its survival time. Then:
\begin{align}
\mathscr{F}^{\mathcal{T}}(\epsilon_t) := \left\{ \epsilon_j \in \epsilon_t : \mathcal{T}_j \geq \bar{\mathcal{T}} \right\} \label{eq:life_filter}
\end{align}
This removes all events corresponding to orders with a lifetime below threshold \( \bar{\mathcal{T}} \).
\end{tcolorbox}

Lifetime-based filtering removes orders that are ephemeral but does not address the behaviour of frequently modified orders that remain in the LOB. Orders with high modification counts may not reflect a genuine intent to trade. We discard such frequently modified orders by applying a second filter based on the number of post-submission modifications.

\begin{tcolorbox}[colframe=black!60, colback=white!97!gray, boxrule=0.4pt, arc=2pt, left=4pt, right=4pt, top=4pt, bottom=4pt, enhanced, sharp corners=south]
\textbf{Definition: Modification Count Filter, $\mathscr{F}^{M}(\epsilon_t)$}

\smallskip
\noindent
Let \( M_j \) denote the number of modifications associated with order ID \( j \). Then:
\begin{align}
\mathscr{F}^{M}(\epsilon_t) := \left\{ \epsilon_j \in \epsilon_t : M_j \leq \bar{M} \right\} \label{eq:mod_filter}
\end{align}
This removes events tied to orders with modification count exceeding threshold \( \bar{M} \).
\end{tcolorbox}

Although the modification count captures the intensity of changes over the lifespan of an order, it does not reflect how concentrated those changes are in time. Orders that undergo rapid, successive modifications just before cancellation may serve to spoof or mislead. The final filtration scheme targets such temporally clustered behavior by requiring a minimum separation between the last two modifications, thereby removing orders that exhibit tightly packed adjustment bursts.

\begin{tcolorbox}[colframe=black!60, colback=white!97!gray, boxrule=0.4pt, arc=2pt, left=4pt, right=4pt, top=4pt, bottom=4pt, enhanced, sharp corners=south]
\textbf{Definition: Modification Time Filter, $\mathscr{F}^{\mathcal{M}}(\epsilon_t)$}

\smallskip
\noindent
Let \( \mathcal{M}_j \) be the time between the last two modifications of order \( j \). Then:
\begin{align}
\mathscr{F}^{\mathcal{M}}(\epsilon_t) := \left\{ \epsilon_j \in \epsilon_t : \mathcal{M}_j \geq \bar{\mathcal{M}} \right\} \label{eq:modtime_filter}
\end{align}
This removes all events corresponding to orders with tightly clustered final modifications, i.e., \( \mathcal{M}_j < \bar{\mathcal{M}} \).
\end{tcolorbox}

\paragraph{Summary of Filtration Schemes.}
Each of the three filtration schemes introduced above is applied independently in our analysis, allowing us to isolate specific structural noise. The order lifetime filter removes short-lived orders that are unlikely to represent genuine trading interest. The modification count filter targets orders with frequent quote updates, often indicative of speculative or non-committal behavior. The modification time filter excludes orders with rapid adjustments near the end of their lifetime. By treating these filters separately, we assess their individual contributions to signal stability and directional clarity without confounding their effects.

\subsection{Evaluation Indicators: Order Book Imbalance (OBI and OBI\textsuperscript{(T)})}
To support the interpretation of score behavior across filtration schemes, we employ OBI as a regime-defining indicator. Although not directly used in scoring, OBI provides a tick-level measure of passive liquidity pressure that helps to interpret how structural filters alter the alignment of an established directional signal~\cite{cont2014price} with the short-horizon returns.

We use the Order Book Imbalance (OBI) measure introduced in Equation~\eqref{eq:ofi}, which captures the directional skew in event arrivals over a backward-looking window \( (\tau - h, \tau] \). OBI values close to \(\pm1\) indicate an extreme imbalance and are often associated with the future direction of price movement. We compute OBI from the LOB reconstructed after each filtration scheme and then segment return-score behavior into distinct imbalance regimes, highlighting where structural filtering yields the greatest directional alignment.

Although order-based OBI provides a real-time view of latent book pressure, it is susceptible to noise from fleeting orders and strategic placements. To construct an execution-based benchmark, where the noise induced by ambiguous intent is mitigated, we introduce an alternative imbalance measure derived solely from signed trade activity. This trade-based OBI~\cite{nittur2025tradeobi} captures the realised directional pressure by contrasting buyer-initiated and seller-initiated trades within each window. This resulting signal is immune to LOB flicker induced by orders and better reflects market commitment. 

\begin{tcolorbox}[colframe=black!60, colback=white!97!gray, boxrule=0.4pt, arc=2pt, left=4pt, right=4pt, top=4pt, bottom=4pt, enhanced, sharp corners=south]
\textbf{Definition: Order Book Imbalance by Trades, $OBI^{(T)}$}

\smallskip
\noindent
Let \( N_{[\tau - h, \tau)}^{(b, T)} \) and \( N_{[\tau - h, \tau)}^{(s, T)} \) denote the total number of trades initiated by buyers and sellers, respectively, within window \( [\tau - h, \tau) \). Then:
\begin{align}
\text{OBI}^{(T)}(\tau) := \frac{\Delta N_{[\tau - h, \tau)}^{(b,T)} - \Delta N_{[\tau - h, \tau)}^{(s,T)}}{\Delta N_{[\tau - h, \tau)}^{(b,T)} + \Delta N_{[\tau - h, \tau)}^{(s,T)}}
\end{align}
\end{tcolorbox}

Trade-based OBI differs from order-derived imbalance. The OBI is derived from the relative difference in the counts of buy and sell orders, capturing latent liquidity. The trade-based \( \text{OBI}^{(T)} \), is based exclusively on realised executions. Specifically, we follow the methodology proposed in~\cite{nittur2025tradeobi}, where each trade is signed using a tick-based classification rule and then imbalance is computed as the net signed trade count within each evaluation window. This construction eliminates noise from flickering quotes, non-executed strategic placements, and ephemeral cancellations. 

\subsection{Scoring Functionals}
\label{subsec:score_functionals}

To evaluate the directional informativeness of filtered OBI, we use a sequence of scoring functionals that quantify the strength and structure of the association between OBI and returns. The functionals are ordered by increasing structural sophistication. We begin with Pearson correlation, which measures the linear association between instantaneous OBI and returns. We then examine the linear association between classified regimes of OBI and returns, which reveal directional alignment. Finally, we fit a Hawkes process to the classified order flow, using the same classification scheme as defined to examine the linear association, and then examine the causal relationship between OBI and returns regimes by analysing Hawkes kernel norms. This layered construction allows us to distinguish linear association between the \emph{values} of OBI and returns from regime-specific \emph{directional} alignment, with statically defined lags, and with dynamic memory structures using Hawkes kernel norms.

\paragraph{(1) Contemporaneous Correlation Score}

The first diagnostic layer assesses the contemporaneous Pearson correlation between the OBI signal values and the traded returns, evaluated over fixed-length time windows. This correlation coefficient provides a measure of linear association and serves as a baseline indicator for signed pressure in the order flow and price changes. It does not capture temporal directionality or distinguish causal from coincidental alignment. We note that although it is sensitive to short-lived volatility in both series, contemporaneous correlation provides an interpretable scalar summary.

\begin{tcolorbox}[colframe=black!60, colback=white!97!gray, boxrule=0.4pt, arc=2pt, left=4pt, right=4pt, top=4pt, bottom=4pt, enhanced, sharp corners=south]
\textbf{Definition: Contemporaneous Correlation Score, $\mathcal{S}^{\rho}(\mathscr{F}, \tau)$}

\smallskip
Let \( X = \text{OBI}({\tau}) \), under filtration $\mathscr{F}^{(\cdot)}$ and \( Y = \tilde{r}_{\tau} \) denote the averaged OBI and return over window \( [\tau - h, \tau) \). Then:

\begin{align}
\mathcal{S}^{\rho}(\tau) := \rho(X, Y) = \frac{\mathbb{E}[(X - \mathbb{E}[X])(Y - \mathbb{E}[Y])]}{\sqrt{\mathbb{V}[X] \cdot \mathbb{V}[Y]}}
\end{align}

where expectations are computed over a rolling ensemble of evaluation windows.
\end{tcolorbox}

\paragraph{(2) Explanatory Power under Discretized Regimes}

To sharpen our investigation on the directional alignment between OBI and returns, we classify both the OBI and returns into discrete regimes. Starting from the tick-level OBI series and the realized return series defined above, we fix symmetric thresholds on their empirical support. The OBI grid partitions the support into nine ordered regimes, from strong negative to strong positive imbalance. The return grid partitions the support into three regimes: positive returns, negative returns, and a small neutral band around zero.

For each evaluation window $[\tau - h, \tau)$, we assign every tick to an OBI regime and a return regime using these thresholds. We then construct two regime-count vectors. The first, $Q_\tau \in \mathbb{R}^9$, records how many times each of the nine OBI regimes appears in the window. The second, $R_\tau \in \mathbb{R}^3$, records how often each of the three return regimes occurs. This produces an aligned time series of regime-count vectors $\{Q_\tau\}$ and $\{R_\tau\}$ throughout the sample.

Within this regime space, we measure directional alignment in two complementary ways. First, for each window $\tau$ we compute a matrix of Pearson correlation coefficients between OBI regime counts and return regime counts, denoted by $\rho_\tau \in \mathbb{R}^{9 \times 3}$. We then average these matrices over $\tau$ to obtain an aggregate correlation matrix. From this matrix we retain only the entries corresponding to directionally aligned regimes, where positive OBI regimes pair with the positive return regime and negative OBI regimes pair with the negative return regime. The directional correlation score increases when correlation becomes stronger and more concentrated in these cells.

Second, we measure explanatory power through a multivariate linear regression of return regime counts on OBI regime counts. For each window, we regress $R_\tau$ on $Q_\tau$ and compute the coefficient of determination $R^2(\tau)$. Aggregating these values over all windows yields a regression-based regime score that summarizes how well the composition of OBI regimes explains the composition of return regimes.

To account for serial dependence, we implement residualised counts. For each component of the OBI regime-count series and for each component of the return regime-count series, we fit a univariate ARMA model and extract residuals. We then recompute both the directional correlation score and the regression-based regime score using these residual series. This isolates the association between those parts of the OBI and returns that are not explained by their own past. Finally, we repeat the entire procedure on a grid of positive lags. For each lag $\ell$, we shift the returns regime-count series forward by $\ell$ while keeping the OBI regime-count series fixed and recompute both scores. This lagged analysis tests whether filtered OBI regimes carry directional information for future return regimes rather than simply co-moving contemporaneously.

\begin{tcolorbox}[colframe=black!60, colback=white!97!gray, boxrule=0.4pt, arc=2pt,
left=4pt, right=4pt, top=4pt, bottom=4pt, enhanced, sharp corners=south]
\textbf{Definition: Directional Correlation Regime Score.}

\smallskip
\noindent
For each window \( \tau \in \mathcal{T} \), let \( Q_\tau \in \mathbb{R}^{9} \) and \( R_\tau \in \mathbb{R}^{3} \) denote the OBI and return regime--count vectors. We form the window--specific Pearson correlation matrix
\[
\rho_\tau \in \mathbb{R}^{3 \times 9},
\]
whose entry \( \rho_{\tau,ij} \) is the correlation between the count of return regime \( i \) and the count of OBI regime \( j \) in window \( \tau \). The averaged correlation matrix over all \( B \) evaluation windows is
\begin{align}
\bar{\rho}(\mathscr{F})
:=
\frac{1}{B} \sum_{\tau \in \mathcal{T}} \rho_\tau(\mathscr{F}),
\end{align}
where the dependence on the filtration scheme \( \mathscr{F} \) is made explicit.

\smallskip
\noindent
Given a directional weight matrix
\(
W = (W_{ij})_{i=0,\dots,n-1;\; j=0,\dots,m-1}
\)
defined below, the directional correlation score is defined as a signed, regime--weighted aggregate of the average correlation matrix, which is positive when directionally aligned regime pairs dominate and negative when misaligned structure dominates:
\begin{align}
\mathcal{S}^{\rho}_{\text{OBI}\rightarrow \tilde{r}}(\mathscr{F})
:=
\sum_{i=0}^{n-1} \sum_{j=0}^{m-1}
W_{ij}\,\bar{\rho}_{ij}(\mathscr{F}).
\end{align}
\end{tcolorbox}

\begin{tcolorbox}[colframe=black!60, colback=white!97!gray, boxrule=0.4pt, arc=2pt,
left=4pt, right=4pt, top=4pt, bottom=4pt, enhanced, sharp corners=south]
\textbf{Directional Mask.}

\smallskip
\noindent
To emphasise directionally aligned regime pairs while retaining the sign and overall structure of \( \bar{\rho}(\mathscr{F}) \), we introduce a smooth weight matrix
\begin{align}
W = (W_{ij})_{i=0,\dots,n-1;\; j=0,\dots,m-1},
\end{align}
with entries
\begin{align}
W_{ij} = 1 + b_i a_j,
\end{align}
and row and column factors
\begin{align}
a_j = j - \frac{m-1}{2},
\qquad
b_i = -\gamma \Bigl(i - \frac{n-1}{2}\Bigr),
\end{align}
for
\[
i = 0,\dots,n-1,\;\; j = 0,\dots,m-1,
\]
and a fixed slope parameter \( \gamma > 0 \) (in our implementation \( \gamma = 0.2 \)). Indices \( j \) enumerate OBI regimes from strongly negative to strongly positive, and indices \( i \) enumerate return regimes from strongly positive to strongly negative. This construction yields weights larger than one for cells along the anti--diagonal corresponding to directionally aligned OBI and return regimes, weights close to one near the neutral--neutral centre, and weights below one for misaligned pairs.

\smallskip
\noindent
The weight matrix thus acts as a smooth directional mask: it preserves the sign and full pattern of \( \bar{\rho}(\mathscr{F}) \) while upweighting directionally aligned OBI--return regime pairs and down--weighting misaligned pairs.
\end{tcolorbox}

\begin{tcolorbox}[colframe=black!60, colback=white!97!gray, boxrule=0.4pt, arc=2pt,
left=4pt, right=4pt, top=4pt, bottom=4pt, enhanced, sharp corners=south]
\textbf{Definition: Regression–Based Regime Score.}

\smallskip
\noindent
For each window \( \tau \in \mathcal{T} \), we fit a multivariate linear regression of return regime counts on OBI regime counts,
\begin{align}
R_\tau = \beta Q_\tau + \varepsilon_\tau,
\end{align}
via ordinary least squares and compute the coefficient of determination \( R^2(\tau) \). The regression–based regime score aggregates explanatory power across windows:
\begin{align}
\mathcal{S}^{\mathcal{R}}_{\text{OBI} \rightarrow r}(\mathscr{F})
:=
\sum_{\tau \in \mathcal{T}} R^2(\tau).
\end{align}
In the residualised variant, the vectors \( Q_\tau \) and \( R_\tau \) are replaced by residual regime–count vectors obtained from univariate ARMA models fitted to each component of the original regime–count processes.
\end{tcolorbox}

To visualise the regime-level structure behind these scores, we use Figure~\ref{fig:point_process_regimes}. The figure shows a schematic $9 \times 3$ grid whose horizontal axis corresponds to OBI regimes and whose vertical axis corresponds to the return regimes for a generic lag $\ell$. Cells in the top-right and bottom-left corners represent directionally aligned pairs, where strongly positive (negative) OBI regimes co-occur with positive (negative) return regimes.

\begin{figure}[H]
\centering
\begin{tikzpicture}[x=0.8cm,y=0.8cm]

% helper macro: one 9x3 panel at vertical offset #1 with title #2
\newcommand{\RegimePanel}[2]{%
  \begin{scope}[shift={(0,#1)}]
    % light background for all cells
    \foreach \x in {0,...,8}{
      \foreach \y in {0,...,2}{
        \fill[blue!10] (\x,\y) rectangle ++(1,1);
      }
    }
    % highlight directional cells: bottom-left, top-right, and centre
    \foreach \x in {0,1,2,3}{
      \fill[blue!60] (\x,0) rectangle ++(1,1);
    }
    \foreach \x in {5,6,7,8}{
      \fill[blue!60] (\x,2) rectangle ++(1,1);
    }
    \fill[blue!40] (4,1) rectangle ++(1,1); % central neutral regime

    % grid lines
    \draw[thin] (0,0) rectangle (9,3);
    \foreach \x in {1,...,8}{\draw[thin] (\x,0) -- (\x,3);}
    \foreach \y in {1,2}{\draw[thin] (0,\y) -- (9,\y);}

    % x-axis tick labels: OBI regimes -4,...,4
    \foreach \x/\lab in {0/-4,1/-3,2/-2,3/-1,4/0,5/1,6/2,7/3,8/4}{
      \node[below] at (\x+0.5,-0.2) {\scriptsize $\lab$};
    }
    \node[below] at (4.5,-0.9) {\scriptsize OBI regimes};

    % y-axis tick labels: return regimes -1,0,1
    \foreach \y/\lab in {0/-1,1/0,2/+1}{
      \node[left] at (-0.2,\y+0.5) {\scriptsize $\lab$};
    }
    \node[rotate=90] at (-1.1,1.5) {\scriptsize Return regimes};

    % panel title
    \node[above,font=\scriptsize] at (4.5,3.4) {Spectrum for lagged $\ell$ s};
  \end{scope}
}

% three panels: 1s, 10s, 50s
\RegimePanel{0}{1 s}
%\RegimePanel{-5}{10 s}
%\RegimePanel{-10}{50 s}

\end{tikzpicture}
\caption{\footnotesize Schematic illustration of the $9 \times 3$ regime correlation structure between OBI regimes (horizontal axis) and return regimes (vertical axis) for three lags. Darker cells indicate directionally aligned pairs.}
\label{fig:point_process_regimes}
\end{figure}

\begin{tcolorbox}[colframe=black!40!white, colback=gray!3, boxrule=0.3pt]
\textbf{Modeling Remark.}

Regime binning is performed symmetrically around zero, with uniformly spaced intervals across $[-1, 1]$ and a central neutral regime for both OBI and returns. The directional correlation score $\mathcal{S}^{\rho}_{\text{OBI} \rightarrow r}(\mathscr{F})$ focuses on OBI--return regime pairs that are directionally aligned and downweights off-regime associations by construction. To evaluate short-term predictive alignment, we recompute both $\mathcal{S}^{\rho}_{\text{OBI} \rightarrow r}(\mathscr{F})$ and $\mathcal{S}^{\mathcal{R}}_{\text{OBI} \rightarrow r}(\mathscr{F})$ across a grid of positive lags $\ell$, where OBI regimes at time $\tau$ are paired with return regimes at $\tau + \ell$. To remove spurious effects from autocorrelation, we repeat the analysis using ARMA residuals for both regime-count processes. Reporting both raw and residualised scores highlights how robust directional association is to serial dependence under each filtration scheme.
\end{tcolorbox}

\paragraph{(3) Hawkes Excitation Norms}

The final layer in our diagnostic ladder uses a multivariate Hawkes process to capture causal excitation in event time. A Hawkes process is a point process whose conditional intensity increases after past events and then decays over time \cite{hawkes1971spectra}. In the multivariate case, each component can excite both itself and the others through a matrix of kernels, which makes the model well suited to high-frequency order flow where trades and quote changes arrive in bursts and past activity feeds back into future arrivals \cite{bacry2015hawkes}. In financial markets, Hawkes models have been used to quantify \emph{reflexivity}, that is, the degree to which price changes are endogenous to past order flow and price moves \cite{hardiman2013reflexivity}.

In our setting, we treat regime-labeled OBI and return events as components of a joint Hawkes process. For each tick, OBI is assigned to one of nine regimes and returns to one of three regimes, as described in the previous subsection. We then construct a multivariate counting process on a common timeline, where some components represent OBI regime events and the remaining components represent return regime events. The Hawkes specification links these components through a matrix of excitation kernels that encode how the arrival of an event in a given OBI regime affects the future intensity of return events in each return regime.

% Box 1: Hawkes process and integrated kernel
\begin{tcolorbox}[colframe=black!60, colback=white!97!gray, boxrule=0.4pt, arc=2pt,
left=4pt, right=4pt, top=4pt, bottom=4pt, enhanced, sharp corners=south]
\textbf{Definition: Multivariate Hawkes Process and Integrated Kernel.}

\smallskip
\noindent
Let \( N(t) = \big(N^{(1)}(t),\dots,N^{(d)}(t)\big)^\top \) denote a \(d\)-dimensional counting process representing all OBI and return regimes. A multivariate Hawkes process has conditional intensity vector \( \lambda(t) = \big(\lambda^{(1)}(t),\dots,\lambda^{(d)}(t)\big)^\top \) of the form
\begin{align}
\lambda^{(i)}(t)
=
\mu^{(i)}
+
\sum_{j=1}^d \int_{0}^{t} \phi_{ij}(t-s)\,\mathrm{d}N^{(j)}(s),
\end{align}
\[
\quad i = 1,\dots,d,
\]
where \( \mu^{(i)} \) is a baseline intensity and \( \phi_{ij}(\cdot) \) is the excitation kernel from component \( j \) to component \( i \). Following \cite{nittur2025tradeobi}, we use a sum-of-exponentials parameterization,
\begin{align}
\phi_{ij}(u)
=
\sum_{m=1}^{M} \alpha_{ij}^{(m)} \beta_m \mathrm{e}^{-\beta_m u} \mathbf{1}_{\{u>0\}},
\end{align}
with common decay rates \( \{\beta_m\}_{m=1}^M \) and component-specific amplitudes \( \{\alpha_{ij}^{(m)}\} \).

\smallskip
\noindent
From the estimated kernels we construct the integrated excitation matrix
\begin{align}
\Phi
=
\left[
\int_0^{\infty} \phi_{ij}(u)\,\mathrm{d}u
\right]_{i,j=1}^d,
\end{align}
whose entry \( \Phi_{ij} \) measures the expected number of offspring events of type \( i \) triggered by a single event of type \( j \).
\end{tcolorbox}

In line with the diagnostic ladder, we focus on cross-excitation from OBI regimes to return regimes. Let \( \Phi_{\text{OBI}\rightarrow r} \) denote the submatrix of \( \Phi \) that maps the nine OBI regimes to the three return regimes. A large and diagonally concentrated \( \Phi_{\text{OBI}\rightarrow r} \) indicates that strongly positive (negative) OBI regimes tend to trigger positive (negative) return regimes after accounting for the full event-time history. To summarise this structure in a single scalar, we define the Hawkes excitation norm score as the Euclidean norm of the entries of \( \Phi_{\text{OBI}\rightarrow r} \) corresponding to directionally aligned regime pairs.
% Box 2: Excitation norm score
\begin{tcolorbox}[colframe=black!60, colback=white!97!gray, boxrule=0.4pt, arc=2pt,
left=4pt, right=4pt, top=4pt, bottom=4pt, enhanced, sharp corners=south]
\textbf{Definition: Hawkes Excitation Norm Score.}
\smallskip
\noindent
\\
Let \( \mathcal{A} \) index OBI–return regime pairs with matching sign (strongly negative OBI and negative returns, neutral OBI and small returns, strongly positive OBI and positive returns). The scoring functional is
\begin{align}
\mathcal{S}^{\phi}_{\text{OBI}\rightarrow r}(\mathscr{F})
:=
\sum_{(i,j)\in \mathcal{A}}
\Phi_{\text{OBI}\rightarrow r}^{(i,j)}
,
\end{align}
which increases when a filtration scheme \( \mathscr{F} \) yields a clearer, more directional excitation pattern from imbalance regimes to return regimes.
\end{tcolorbox}

\begin{tcolorbox}[colframe=black!40!white, colback=gray!3, boxrule=0.3pt]
\textbf{Estimation Remark.}
Although the multivariate Hawkes specification entails a high parameter count (a \(13\times13\) kernel with multiple exponential components), the estimation sample is of commensurate scale at the event level. The large number of observed LOB events provides sufficient information to estimate the excitation structure and to assess filtration effects without relying on aggressive regularization.
\end{tcolorbox}

\paragraph{(4) OBI by trades, $\text{OBI}^{(T)}$}
In addition to the imbalance signals derived from the LOB state under the filtration schemes above, we construct a trade-based imbalance signal that is directly tied to executed transactions. For each evaluation window, we compute the trade-based imbalance $\text{OBI}^{(T)}$ from the net signed trade count, using the tick-based trade classification and windowed construction defined in the preceding subsection.

To make this object comparable to our structural filters, we apply each of the lifetime, modification count, and modification time filters to the parent orders that resulted in the trades, identified by the order ID present in the tick-by-tick data for every trade. For a given filtration scheme $\mathscr{F}$, we retain a trade in the window only if its parent order satisfies the corresponding threshold for that single filter. Trades whose parent orders fail the lifetime threshold under $\mathscr{F}^{\mathcal{T}}$, the modification count threshold under $\mathscr{F}^{M}$, or the modification time threshold under $\mathscr{F}^{\mathcal{M}}$ are discarded from the analysis for that filter. The resulting trade-based signals, which we denote by $\text{OBI}^{(T,\mathscr{F})}$, therefore reflect the realised directional pressure carried by orders that would also have survived the corresponding structural filter in the book.

Compared to standard book-derived OBI, which can be influenced by fleeting or unexecuted quotes, $\text{OBI}^{(T)}$~\cite{nittur2025tradeobi} is anchored in executed transactions and is less sensitive to short-lived book noise. In our diagnostic ladder, we use both the unfiltered trade-based signal $\text{OBI}^{(T)}$ and its parent-filtered variants $\text{OBI}^{(T,\mathscr{F})}$ as benchmarks when interpreting the Hawkes excitation scores, particularly the cross-kernel norms from trade-based OBI regimes to future return regimes.

\begin{figure}[H]
\centering
\resizebox{0.95\columnwidth}{!}{%
\begin{tikzpicture}[node distance=1.4cm and 2.2cm, every node/.style={font=\footnotesize}, >=latex]

  % Input block
  \node[draw, rectangle, fill=gray!10, minimum width=2.0cm, minimum height=0.9cm] (input) {Raw OBI \& Return Data};

  % Filter blocks
  \node[draw, rectangle, fill=blue!10, below left=of input, minimum width=2.0cm, minimum height=0.9cm] 
    (filter1) {\shortstack{Lifetime Filter\\ \( \mathscr{F}^{\mathcal{T}} \)}};

  \node[draw, rectangle, fill=blue!10, below=of input, minimum width=2.0cm, minimum height=0.9cm] 
    (filter2) {\shortstack{Modification Count Filter\\ \( \mathscr{F}^{M} \)}};

  \node[draw, rectangle, fill=blue!10, below right=of input, minimum width=2.0cm, minimum height=0.9cm] 
    (filter3) {\shortstack{Modification Time Filter\\ \( \mathscr{F}^{\mathcal{M}} \)}};

  % Score blocks
  \node[draw, rectangle, fill=green!10, below=2.1cm of filter1, minimum width=2.0cm, minimum height=0.9cm] 
    (score1) {\shortstack{Correlation Score\\ \( \mathcal{S}^{\rho} \)}};

  \node[draw, rectangle, fill=green!10, below=2.1cm of filter2, minimum width=2.0cm, minimum height=0.9cm] 
    (score2) {\shortstack{Regime Score\\ \( \mathcal{S}^{\mathcal{R}} \)}};

  \node[draw, rectangle, fill=green!10, below=2.1cm of filter3, minimum width=2.0cm, minimum height=0.9cm] 
    (score3) {\shortstack{Hawkes Score\\ \( \mathcal{S}^{\phi} \)}};

  % Arrows from input to filters
  \draw[->, thick] (input.south west) -- (filter1.north);
  \draw[->, thick] (input.south) -- (filter2.north);
  \draw[->, thick] (input.south east) -- (filter3.north);

  % Arrows from filter1 to all scores
  \draw[->, thick] (filter1.south) -- (score1.north);
  \draw[->, thick] (filter1.south) to (score2.north west);
  \draw[->, thick] (filter1.south) to (score3.north);

  % Arrows from filter2 to all scores
  \draw[->, thick] (filter2.south) -- (score2.north);
  \draw[->, thick] (filter2.south) to (score1.north east);
  \draw[->, thick] (filter2.south) to (score3.north west);

  % Arrows from filter3 to all scores
  \draw[->, thick] (filter3.south) -- (score3.north);
  \draw[->, thick] (filter3.south) to (score2.north east);
  \draw[->, thick] (filter3.south) to (score1.north);

\end{tikzpicture}%
}
\caption{\footnotesize Schematic overview of the filtration and scoring pipeline. Raw order book imbalance and return streams are processed through three distinct filtering schemes. The resulting filtered data are then evaluated using correlation, regime-based, and Hawkes excitation scoring functionals. This layered structure allows a modular assessment of signal quality under different microstructure filters.}
\label{fig:filtering_scoring_pipeline}
\end{figure}

\paragraph{Summary of Scoring Functionals.}
Figure~\ref{fig:filtering_scoring_pipeline} shows the hierarchical schema of our scoring framework. The layered structure offers a coherent diagnostic ladder for evaluating directional informativeness. The first layer assesses value-level co-movement using raw Pearson correlation. The second captures structural alignment through regime-discretized counts and explanatory regression. The third layer models causal excitation using point process dynamics. Together, these layers reveal how filtration clarifies directional signals by reducing noise, emphasizing regime-level structure, and uncovering causal influence.

To maintain strict comparability across filtration schemes, all scoring functionals are computed using identical evaluation windows and normalization logic. For each variant, we reconstruct the limit order book from the filtered event stream, extract directional signals using the same conventions, and apply the scoring pipeline without model-specific tuning. This ensures that any observed differences in association strength can be attributed solely to structural filtering, without confounding effects from reparameterization or inconsistent return handling.

\section{Data}
\label{sec:data}

Our empirical study is based on tick-by-tick data for BANKNIFTY futures and selected equities sourced from the Indian National Stock Exchange (NSE). Each tick records a discrete event involving one of four categories: new order submission, order modification, order cancellation, or trade execution. Every order is uniquely identified by an Order ID (OID), which remains associated with all its subsequent modification events until the order is either cancelled or executed.

\smallskip
Each tick includes a market snapshot comprising the top five levels of the limit order book (LOB), including bid/ask prices and corresponding standing volumes. Importantly, the dataset is \emph{event-driven}: only those ticks that induce a change in the top-5 levels of the book are recorded. As a result, periods of inactivity or activity beyond the top 5 levels are omitted from the stream, yielding a compact, latency-sensitive representation of market activity.

\smallskip
To analyze how filtering affects the directional information content of the order flow, we reconstruct alternate versions of the limit order book under different filtration schemes. The filtering process is implemented as follows:

\begin{enumerate}
    \item For each filtration scheme \( \mathscr{F}^{(\cdot)} \), we construct a list of Order IDs (OIDs) that are to be excluded from contributing to the displayed book.
    \item During reconstruction, any new or modified order matching an excluded OID is ignored, except in the case of trade execution. If a filtered OID executes against a resting order, its trade tick is retained to preserve execution consistency and prevent structural imbalance in the reconstructed book.
\end{enumerate}

This protocol ensures that downstream scoring functionals operate over books that remain structurally valid and comparable across filtration schemes.

\section{Experimental Setup}
\label{sec:experiment}

We work with tick-level data for BANKNIFTY January 2023 index futures traded on the National Stock Exchange of India. The sample consists of three trading days: 2 January, 13 January and 23 January 2023, chosen to span the early, middle and late parts of the expiry cycle and to exhibit varying market activity. Each event record contains the server timestamp, order identifier, event type (NEW, MODIFY, CANCEL, TRADE), side (bid or ask), price and quantity. Using these event streams, we reconstruct the limit order book.

Evaluation is performed on overlapping time windows. For each anchor time \( \tau \), we define an evaluation window \([ \tau, \tau + h )\) of length \( h = 10 \) seconds, over which we compute an imbalance signal, and a subsequent forecast window \((\tau, \tau + \xi]\) of length \( \xi = 1 \) second, over which we compute the realised traded return \( \tilde{r} \). Anchor times are spaced every 15 seconds during the main continuous trading session (09:20--15:25), yielding a large number of non-overlapping forecast windows.

On the event stream \( \epsilon_t \), we apply three filtration schemes defined in Section~\ref{subsec:filters}. The lifetime filter \( \mathscr{F}^{\mathcal{T}} \) retains only events associated with orders whose total survival time \( \mathcal{T}_j \) exceeds a threshold \( \bar{\mathcal{T}} \). The modification-count filter \( \mathscr{F}^{M} \) retains events for orders with modification counts \( M_j \) below a threshold \( \bar{M} \). The modification-time filter \( \mathscr{F}^{\mathcal{M}} \) retains events where the time between order entry and the first modification, \( \mathcal{M}_j \), exceeds a threshold \( \bar{\mathcal{M}} \). Each scheme can be implemented in real time using only observable order histories.

Filtration is parametrised by simple threshold grids. For lifetime we consider \( \bar{\mathcal{T}} = \ 100 \) milliseconds, for modification time we appy \( \bar{\mathcal{M}} = \ 50 \) milliseconds, and for modification count \( \bar{M} = \ 3 \) with multiple lags being used to determine correlation and regime correlation scores. For each combination of date, threshold and lag, we apply one filtration scheme at a time to the same underlying event stream.

To give a sense of scale, a single BANKNIFTY session typically produces on the order of \(10^8\) events. On 23 January 2023, for example, the unfiltered book reconstruction uses 83{,}547{,}534 ticks. Applying the lifetime filter removes 5{,}928{,}152 order identifiers and yields 51{,}023{,}701 ticks in the filtered stream. The modification-count filter removes 5{,}416{,}670 order identifiers and leaves 72{,}228{,}161 ticks, while the modification-time filter removes 168{,}058 order identifiers and leaves 47{,}531{,}611 ticks. Thus each structural filter removes a sizeable set of short-lived or heavily modified orders but still leaves tens of millions of events; any change in directional diagnostics reflects altered signal content rather than a collapse in sample size.

From each filtered event stream we construct order book imbalance (OBI) over \([ \tau, \tau + h )\) and pair it with the realised return \( \tilde{r} \) over the forecast window. These LOB-based imbalance signals are evaluated using the full diagnostic ladder described in Section~\ref{sec:methodology}: value-level correlations, discretised regime association and Hawkes-based excitation norms. In parallel, we construct a trade-based imbalance signal \( \text{OBI}^{(T)} \) using signed trade events linked to their parent orders. For this trade-based signal we focus on Hawkes-based excitation diagnostics, since any value-level correlation or regime association between trade-based imbalance and short-horizon returns is mechanically strong and less informative for our purposes.

\begin{table}[H]
\centering
\caption{Event-level impact of filtration on BANKNIFTY January futures, 23 January 2023 (T--2 before expiry)}
\label{tab:event_level_filtration}
\resizebox{0.9\columnwidth}{!}{%
\begin{tabular}{|l|r|r|}
\hline
\textbf{Filter} & \textbf{Filtered orders count} & \textbf{Final ticks} \\
\hline
Default book (no filtration)         & 0        & 83{,}547{,}534 \\
Lifetime filter \( \mathscr{F}^{\mathcal{T}} \)  & 5{,}928{,}152 & 51{,}023{,}701 \\
Mod-count filter \( \mathscr{F}^{M} \)          & 5{,}416{,}670 & 72{,}228{,}161 \\
Mod-time filter \( \mathscr{F}^{\mathcal{M}} \) &   168{,}058  & 47{,}531{,}611 \\
\hline
\end{tabular}
}
\end{table}

These counts highlight that filtration is a non-trivial design problem at scale. A single BANKNIFTY session produces on the order of \(10^8\) events, and raw OBI on this stream mixes genuine directional pressure with very short-lived and repeatedly modified orders. Structural filters \( \mathscr{F}^{\mathcal{T}}, \mathscr{F}^{M}, \mathscr{F}^{\mathcal{M}} \) each remove on the order of \(10^6\) order identifiers while leaving tens of millions of events in the reconstructed book. Notably, the modification-time filter \( \mathscr{F}^{\mathcal{M}} \) removes only 168{,}058 order IDs yet eliminates approximately 36 million ticks relative to the unfiltered stream, more than the lifetime-based filter. This indicates that a relatively small subset of orders generates a disproportionately large volume of rapid modification activity without leading to execution. In light of this structure, we investigate whether excluding this transient component of the order flow improves the directional content of imbalance-based signals, rather than merely reducing the sample size.

\section{Results}
\label{sec:results}

\subsection{Overview}

We find a sharp contrast between order-based and trade-based imbalance under structural filtration. For imbalance computed from the standing book, lifetime, modification-count and modification-time filters produce only small and unstable changes in directional association with short-horizon returns across all three diagnostic layers. For trade-based imbalance \( \text{OBI}^{(T)} \), the same filters consistently raise Hawkes excitation from imbalance regimes to future return regimes on all three sample days. The subsections below first summarise the weak and heterogeneous effects for LOB-based OBI, and then document the much clearer Hawkes-based gains for trade-filtered \( \text{OBI}^{(T)} \).

Unless otherwise stated, results are reported by filtration scheme and trading day and then summarised by averages across the three dates. The unfiltered series (UF) provides the baseline against which the lifetime (LF), modification-count (MF) and modification-time (MTF) filters are compared.

\subsection{LOB-based Imbalance: Small and Heterogeneous Effects}

For imbalance constructed from standing book depth, all three diagnostics move only slightly under filtration and no filter dominates the unfiltered baseline. Table~\ref{table:R3-summary-final} reports the average values of the diagnostics for LOB-based OBI across all days and filter thresholds. At the correlation layer, all scores are small in absolute terms. The average Pearson correlation score \( \mathcal{S}^{\rho}(\mathscr{F}) \) between unfiltered OBI and subsequent returns is of the order of \(10^{-2}\), and filtration shifts this baseline only marginally. The modification-time filter (MTF) produces the largest relative increase, moving the average correlation from 0.01018 (UF) to 0.01133, while the lifetime filter (LF) leaves it essentially unchanged and the modification-count filter (MF) lowers it. Across individual days and for varying horizons (reported in Appendix Tables~\ref{tab:appendix_corr_off}, \ref{tab:appendix_corr_on}, \ref{tab:appendix_regime_off}, \ref{tab:appendix_regime_on} and  \ref{tab:appendix_hawkes_obi}), some filter and lag combinations show small gains and others show small losses, with no stable pattern.

The regime-based diagnostics also show modest and mixed changes. The average difference between imbalance regimes and subsequent return regimes is smallest under filtration by modification time, with filtration on order lifetime delivering a mild improvement relative to the unfiltered OBI and filtration by modification count slightly under-performing. The regression-based regime scores \( \mathcal{S}^{\mathcal{R}}_{\text{OBI} \rightarrow r}(\mathscr{F}) \) for return regimes on imbalance regimes are almost identical across filters. Once autocorrelation in the return process is taken into account, the unfiltered and filtered series achieve very similar explanatory power in regime space.

Hawkes-based excitation norms \( \mathcal{S}^{\phi}_{\text{OBI} \rightarrow r}(\mathscr{F}) \) for LOB-based OBI behave in a similar way. On average, filtration by modification time yields the highest cross-kernel norm from imbalance regimes to return regimes, followed by filtration on modification count, with filtration by order lifetime almost indistinguishable from the unfiltered OBI. The differences are small in relative terms, and the ranking of filters varies across days. Taken together, these diagnostics indicate that for imbalance constructed from standing book depth, simple structural filters on lifetimes and modification behaviour do not deliver a robust and stable improvement in directional association, both causal and linear, with short-horizon returns.

\begin{table}[H]
\centering
\label{table:R3-summary-final}
\resizebox{0.9\columnwidth}{!}{%
\begin{tabular}{|l|c|c|c|c|}
\hline
\textbf{Filter Type}
& \textbf{\( \mathcal{S}^{\rho}(\mathscr{F}) \)}
& \textbf{\( \mathcal{S}^{\rho}_{\text{OBI} \rightarrow r}(\mathscr{F}) \)}
& \textbf{\( \mathcal{S}^{\mathcal{R}}_{\text{OBI} \rightarrow r}(\mathscr{F}) \)}
& \textbf{\( \mathcal{S}^{\phi}_{\text{OBI} \rightarrow r}(\mathscr{F}) \)} \\ \hline
Unfiltered (UF)  & 0.01018 & -4.00 & 8.43 & 8.9292 \\ \hline
Lifetime (LF)    & 0.01011 & -3.41 & 8.39 & 8.9287 \\ \hline
Mod Count (MF)   & 0.00826 & -4.44 & 8.42 & 9.0048 \\ \hline
Mod-Time (MTF)   & 0.01133 & -3.39 & 8.37 & 9.1745 \\ \hline
\end{tabular}
}
\caption{Summary of averaged scores across scoring functionals and filtration schemes for LOB-based OBI.\\ \small\textit{Note}: All values are averaged across the three trading days. The Pearson correlation score \( \mathcal{S}^{\rho}(\mathscr{F}) \) and the regime-based scores \( \mathcal{S}^{\rho}_{\text{OBI} \rightarrow r}(\mathscr{F}) \) and \( \mathcal{S}^{\mathcal{R}}_{\text{OBI} \rightarrow r}(\mathscr{F}) \) are computed after removing autocorrelation. The Hawkes excitation score \( \mathcal{S}^{\phi}_{\text{OBI} \rightarrow r}(\mathscr{F}) \) uses the sum-of-exponentials kernel variant.}
\smallskip
\end{table}

\subsection{Trade-Based Imbalance: Stronger Hawkes Excitation Under Filtration}

For trade-based imbalance \( \text{OBI}^{(T)} \), constructed from signed trades, structural filtration on the parent orders linked to the trades produces large and systematic changes in Hawkes excitation norms. We apply the same lifetime-, modification-count-, and modification-time-based filters on the linked parent orders and then compute cross-kernel Hawkes norms from \( \text{OBI}^{(T)} \) regimes to return regimes.

Table~\ref{table:HawkesChunk-TradeOBI} reports the excitation scores by filtration scheme and sample date. On all three days, at least one structural filter produces a substantially higher excitation norm than the unfiltered trade-based benchmark. On 2 January 2023, filtration by order lifetime raises the excitation score from 10.99 (unfiltered $OBI^{(T)}$) to 15.22 ($OBI^{(T)}$ post filtration by order lifetime), with filtration by modification-count and modification-time also above the unfiltered $OBI^{(T)}$. On 13 January 2023, filtration by order lifetime again almost doubles the excitation score relative to unfiltered $OBI^{(T)}$, and both modification-count and modification-time filters deliver intermediate improvements. On 23 January 2023, the day with the highest market activity among the 3 sample days, filtration by modification-time produces the strongest effect, with the excitation score rising from 11.59 (unfiltered $OBI^{(T)}$) to 24.74 under filtration by modification-time, while filtration by order lifetime results in $OBI^{(T)}$ slightly above unfiltered $OBI^{(T)}$ and filtration by modification-count lies slightly below.

Across days, modification-time filtration leads to consistently higher Hawkes excitation norms for \( \text{OBI}^{(T)} \) than for the unfiltered $OBI^{(T)}$, with lifetime and modification-count filtration usually raising excitation as well, although with smaller and less stable gains. The excitation structure becomes more directional in the sense that trade-based imbalance events that survive the filters are more often followed by return regimes of the same sign. This effect is present on every sample date and is much more pronounced than any change observed for LOB-based OBI.

\begin{table}[H]
\centering
\resizebox{\columnwidth}{!}{%
\begin{minipage}{0.25\textwidth}
\centering
\begin{tabular}{|c|c|}
\hline
\multicolumn{2}{|c|}{\textbf{02nd Jan 2023}} \\ \hline
\textbf{Filter} & $\textbf{\( \mathcal{S}^{\phi}_{\text{OBI}^{(T)} \rightarrow r}(\mathscr{F}) \)}$ \\ \hline
UF  & 10.9933 \\ \hline
LF  & 15.2172 \\ \hline
MF  & 13.8234 \\ \hline
MTF & 12.0679 \\ \hline
\end{tabular}
\end{minipage}
%\hspace{0.075\textwidth}
\begin{minipage}{0.25\textwidth}
\centering
\begin{tabular}{|c|c|}
\hline
\multicolumn{2}{|c|}{\textbf{13th Jan 2023}} \\ \hline
\textbf{Filter} & $\textbf{\( \mathcal{S}^{\phi}_{\text{OBI}^{(T)} \rightarrow r}(\mathscr{F}) \)}$ \\ \hline
UF  &  8.3639 \\ \hline
LF  & 15.8630 \\ \hline
MF  & 10.2922 \\ \hline
MTF & 12.7363 \\ \hline
\end{tabular}
\end{minipage}
%\hspace{0.075\textwidth}
\begin{minipage}{0.25\textwidth}
\centering
\begin{tabular}{|c|c|}
\hline
\multicolumn{2}{|c|}{\textbf{23rd Jan 2023}} \\ \hline
\textbf{Filter} & $\textbf{\( \mathcal{S}^{\phi}_{\text{OBI}^{(T)} \rightarrow r}(\mathscr{F}) \)}$ \\ \hline
UF  & 11.5868 \\ \hline
LF  & 12.0834 \\ \hline
MF  & 10.8573 \\ \hline
MTF & 24.7352 \\ \hline
\end{tabular}
\end{minipage}
}
\caption{Hawkes excitation scores $\mathcal{S}^{\phi}(\mathscr{F})$ under a sum-of-exponentials kernel, computed from trade-based imbalance $\text{OBI}^{(T)}$, shown by date and filter type. Filter types: UF = Unfiltered, LF = Lifetime Filter, MF = Modification Count Filter, MTF = Modification-Time Filter.}
\label{table:HawkesChunk-TradeOBI}
\end{table}

\subsection{Summary Interpretation}

For OBI computed from the standing book, lifetime-, modification-count- and modification-time-based filters have only small and heterogeneous effects on value-level correlations, regime-based diagnostics and Hawkes excitation norms. In this setting, simple filters on order lifetimes and modification patterns do not yield a reliable improvement in directional alignment with short-horizon returns.

In contrast, for \( \text{OBI}^{(T)} \) constructed from executed trades and filtered via parent-order lifetimes and modifications, the Hawkes diagnostics show a much clearer pattern. On all three sample days, excitation norms from imbalance regimes to return regimes are systematically higher under lifetime and modification-time filtration than under the unfiltered benchmark, with modification-count filtration showing smaller and less stable gains. Within the limits of our three-day sample, this suggests that trades linked to very short-lived or heavily revised orders contribute relatively little to the event-time directional structure of returns, whereas trades that pass simple structural filters carry a clearer causal imprint. The analysis applied to trade-filtered \( \text{OBI}^{(T)} \) reveals filtration effects that remain largely invisible in the correlation and regime-based summaries for book-based OBI.

\section{Discussion and Conclusion}
\label{sec:discussion}

Our study is anchored by the premise that if a large fraction of low-latency message traffic arises from short-lived or heavily revised orders, then filtering such orders should sharpen the directional association between imbalance and short-horizon returns. The evidence is asymmetric. For imbalance constructed from the standing book, filters based on order lifetimes and modification behaviour have only small, filter-dependent effects on the association with subsequent returns. When the same filters are mapped to the parent orders of executed trades and imbalance is constructed from the signed trade flow, Hawkes-based excitation norms between imbalance and return regimes increase substantially and consistently. Thus, tick-level filtration of all orders does not materially alter book-level imbalance signals, whereas filtration applied to trade-linked orders strengthens the event-time directional alignment captured by Hawkes diagnostics.

This pattern aligns with and qualifies existing results on imbalance and message traffic. Work on order flow imbalance and price impact, such as \cite{cont2014impact}, shows that simple unfiltered measures of net order flow already have substantial explanatory power for short-horizon returns. Our LOB-based findings are consistent with this view: the unfiltered imbalance signal already captures most of the linear and regime-level associations recovered by our diagnostics, and simple structural filters do not generate a qualitatively new signal at the aggregate depth. At the same time, the literature on low-latency trading emphasises the prevalence of fleeting orders and rapid cancellations \cite{hasbrouck2013latency, easley2016, dahlstrom2024cancellations}. Our results suggest that, at the standing depth, much of the noise introduced by such low-persistence orders is already averaged out, so structural filtration alone does not produce a markedly sharper LOB-based signal.

On the most volatile day in our sample, this structure is visible in the raw event data: as shown in Table~\ref{tab:event_level_filtration}, removing $\approx 170000$ parent order identifiers under the modification-time filter deletes $\approx 36$ million individual ticks from the order flow, indicating that a small subset of fast-revising orders generates a disproportionate share of market activity even in a post-surveillance regime. At the same time, the corresponding changes in directional scores for order-based imbalance are modest, whereas the trade-based imbalance constructed from the remaining parent orders shows clearer gains in Hawkes excitation towards return regimes. Our filters are agnostic to trading intent and treat both benign risk-management revisions and strategically noisy activity in the same way. The concentration of high-frequency modifications in a small set of parent orders is compatible with an environment where interacting automated strategies can generate emergent order-book churn around sharp price moves, even when overtly malafide patterns are discouraged at the account level. 

Our analysis is conducted in a market where NSE's ``Persistent Noise Creator'' framework is already in force \cite{nse2021pnc,nse2022pncupdate}. The circulars couple high order-to-trade ratios with frequent, priority-lowering modifications on individual clients and are designed to deter persistent noisy order flow. Our results suggest that, even in this regime, a small subset of modified orders can generate a disproportionate share of low-latency message traffic. When structural filters remove these high-activity parent orders, the directional association between trade-based imbalance and return regimes, as measured by Hawkes cross-excitation norms, changes more visibly than the corresponding association for book-based imbalance. We do not observe trader identities or individual latencies, so we do not evaluate the distribution of technological advantages across participants. However, the concentration of rapid modifications in a small set of orders suggests that, in the presence of increased automation across all classes of market participants, the emergent order flow among these market participants may contribute significantly to the aggregate order flow.

In this sense, our filtration-and-diagnostic ladder is not a replacement for traditional market-quality measures, such as spreads and depth, but rather an additional layer that focuses on how order-based regulations affect the aggregate order flow when characterized by short-horizon returns. Multivariate Hawkes specifications provide one concrete way to summarise this causal association through excitation norms between imbalance and return regimes. In the future, such a methodology could be used to compare alternative surveillance parameters or regulatory regimes while considering their effects on aggregate order flow.

Looking ahead, the proposed diagnostic structure can be embedded in a low-touch optimal-control formulation. In such a setting, the lifetime, modification-count, and inter-modification thresholds can be treated as design parameters, and each parameter choice would induce a filtered imbalance series with associated correlation, regime and Hawkes-based diagnostics as loss variables. An information-based objective, such as the directional association between filtered imbalance and short-horizon realised returns, could then be optimised subject to explicit out-of-sample checks across contracts and days. Formalising this calibration scheme and assessing its robustness is beyond the scope of the present methodological study. However, the illustrative results presented suggest that such a control framework could be a promising direction for future work.

% Generate the bibliography.
\bibliography{references}
\clearpage
\section{Appendix - I: Global Notation Summary}

\begin{table}[H]
\centering
\scriptsize
\caption{Summary of Symbols and Their First Point of Introduction}
\label{tab:global_notation_clean}
\renewcommand{\arraystretch}{1.2}
\begin{tabular}{p{3cm}p{6cm}p{3cm}}
\toprule
\textbf{Symbol} & \textbf{Description} & \textbf{Section} \\
\midrule
\multicolumn{3}{l}{\textit{Market Structure and Events}} \\
$p^{b,i}, p^{a,i}$ & Price at level $i$ on bid / ask side & Data / Filtration \\
$q^{b,i}, q^{a,i}$ & Quantity at level $i$ on bid / ask side & Data / Filtration \\
$T$ & Final time in summation/indexed bounds (not event label) & Methodology \\
$T, C$ & Trade and cancellation event types & Problem Description \\
$\mathcal{E}$ & Set of event types (e.g., trades, cancels) & Problem Description \\
$\mathcal{I}$ & Index set over event dimensions & Methodology \\
\midrule

\multicolumn{3}{l}{\textit{Feature Extraction and Window Definitions}} \\
$\tau$ & Anchor time for evaluation window & Problem Description \\
$h$ & Length of evaluation window & Problem Description \\
$\xi$ & Length of forecast window & Problem Description \\
$[\tau - h, \tau)$ & Evaluation window & Problem Description \\
$(\tau, \tau + \xi]$ & Forecast window & Problem Description \\
$\tilde{r}$ & Realized return over forecast window & Forecast Target \\
$\hat{\tilde{r}}$ & Predicted return over forecast window & Experiment / Results \\
$\mathcal{W}$ & Set of evaluation window anchors $\tau$ & Problem Description \\
$\mathcal{T}(w)$ & Tick history over window $w$ & Data / Preprocessing \\
$\mathcal{B}(w)$ & LOB snapshot at start of window $w$ & Data / Preprocessing \\
\midrule

\multicolumn{3}{l}{\textit{Filtering and Feature Construction}} \\
$\mathscr{F}$ & General filtration scheme applied to event stream & Methodology \\
$\mathcal{F}_{\text{q}}$ & Quote-persistence filter & Methodology \\
$\mathcal{F}_{\text{e}}$ & Event excitation filter (e.g., via Hawkes) & Methodology \\
$\mathcal{G}$ & Optional auxiliary filter for OBI-related features & Methodology \\
$\mathcal{K}(w)$ & Feature extraction map over window $w$ & Methodology \\
\midrule

\multicolumn{3}{l}{\textit{Order Book Imbalance and Related Measures}} \\
$\text{OBI}_{[\tau - h, \tau)}$ & Order Book Imbalance in evaluation window & Methodology \\
$\Delta q^{b,1}, \Delta q^{a,1}$ & Changes in top-level quantities & Methodology \\
\midrule

\multicolumn{3}{l}{\textit{Hawkes Process Components}} \\
$\lambda^i(z)$ & Intensity for event type $i$ in fused space & Methodology \\
$\mu^i$ & Baseline intensity for event type $i$ & Methodology \\
$\phi_{ij}$ & Excitation kernel from $j$ to $i$ & Methodology \\
\midrule

\multicolumn{3}{l}{\textit{Prediction and Evaluation}} \\
$\mathcal{S}(\mathscr{F}, \tilde{r})$ & Scoring functional based on filtration and return & Methodology \\
$\mathcal{S}^{(\rho)}$ & Correlation-based scoring functional & Methodology \\
$\mathcal{S}^{(\phi)}$ & Excitation norm–based scoring functional (Hawkes) & Methodology \\
$\rho$ & Pearson correlation coefficient & Evaluation \\
$L$ & Forecast loss (e.g., absolute or squared error) & Evaluation \\
$\mathbb{I}\{\cdot\}$ & Indicator function & Evaluation \\
\midrule

\multicolumn{3}{l}{\textit{Auxiliary Symbols and Derived Constructs}} \\
$\mathcal{M}$ & Filtered model variant or mask & Results \\
$\mathcal{R}$ & Return regime bin identifier & Evaluation \\
$\mathbf{Q}, \mathbf{R}$ & Feature or target vectors for OBI/returns & Methodology / Results \\
$\hat{Q}, \hat{R}$ & Predicted values of feature or response & Evaluation \\
\bottomrule
\end{tabular}
\end{table}

\clearpage
\section{Appendix - II: detailed scores across filtration methods and scoring functionals.}

This appendix reports the day- and horizon-specific quantities from which the aggregate scores
$\mathcal{S}^{\rho}(\mathscr{F})$, $\mathcal{S}^{\rho}_{\text{OBI}\rightarrow \tilde{r}}(\mathscr{F})$,
$\mathcal{S}^{\mathcal{R}}_{\text{OBI}\rightarrow \tilde{r}}(\mathscr{F})$ and
$\mathcal{S}^{\phi}_{\text{OBI}\rightarrow \tilde{r}}(\mathscr{F})$ in the main text are constructed.
For each trading day $d$ and filtration scheme $\mathscr{F} \in \{\text{UF},\text{LF},\text{MF},\text{MTF}\}$,
we report scores at positive horizons $h$ between imbalance and returns, implemented via evaluation windows
$[\tau,\tau+h)$.

\subsection{Pearson Correlation scores across time horizons}
\begin{table}[H]
    \centering
    \resizebox{\columnwidth}{!}{%
    \begin{tabular}{|l|l|l|l|l|l|l|}
    \hline
    \textbf{Date\_Filter} & \textbf{1s} & \textbf{10s} & \textbf{30s} & \textbf{50s} & \textbf{80s} & \textbf{100s} \\ \hline
    20230102\_UF & 0.02067 & 0.00866 & 0.00271 & 0.00368 & 0.00428 & 0.00377 \\ \hline
    20230102\_MTF & 0.01233 & 0.00236 & 0.00070 & 0.00033 & 0.00590 & 0.00120 \\ \hline
    20230102\_LF & 0.02121 & 0.00945 & 0.00463 & 0.00251 & 0.00550 & 0.00332 \\ \hline
    20230102\_MF & 0.00092 & 0.00371 & -0.00520 & 0.00088 & -0.00059 & -0.00151 \\ \hline
    20230113\_UF & 0.00353 & -0.00503 & 0.00043 & 0.00071 & -0.00171 & -0.00334 \\ \hline
    20230113\_MTF & 0.01248 & 0.00308 & -0.00048 & 0.00275 & 0.00164 & -0.00182 \\ \hline
    20230113\_LF & 0.00379 & -0.00486 & 0.00053 & 0.00116 & -0.00272 & -0.00374 \\ \hline
    20230113\_MF & 0.00123 & 0.00127 & -0.00079 & 0.00247 & 0.00076 & -0.00236 \\ \hline
    20230123\_UF & 0.02327 & -0.00325 & 0.00154 & -0.00044 & 0.00507 & -0.00515 \\ \hline
    20230123\_MTF & 0.03022 & -0.00109 & 0.00225 & 0.00184 & 0.00291 & -0.00802 \\ \hline
    20230123\_LF & 0.02135 & -0.00347 & 0.00198 & -0.00104 & 0.00500 & -0.00524 \\ \hline
    20230123\_MF & 0.02453 & 0.00075 & -0.00009 & 0.00088 & 0.00035 & -0.00494 \\ \hline
    \end{tabular}
    }
    \caption{Pearson correlation magnitudes between order-based imbalance OBI and short-horizon realized returns $\tilde{r}$ for each trading day $d$ and filtration scheme $\mathscr{F}$, at horizons $h \in \{1,10,30,50,80,100\}$ seconds (auto-correlation included). Each entry corresponds to a horizon-specific correlation score $\mathcal{S}^{\rho}_{\text{OBI}\rightarrow \tilde{r}}(\mathscr{F})$. Filter types: UF = Unfiltered, LF = Lifetime Filter, MF = Modification Count Filter, MTF = Modification-Time Filter.}
    \label{tab:appendix_corr_off}
\end{table}

\begin{table}[H]
    \centering
    \resizebox{\columnwidth}{!}{%
    \begin{tabular}{|l|l|l|l|l|l|l|}
    \hline
    \textbf{Date\_Filter} & \textbf{1s} & \textbf{10s} & \textbf{30s} & \textbf{50s} & \textbf{80s} & \textbf{100s} \\ \hline
    20230102\_UF & 0.01192 & 0.00656 & 0.00151 & 0.00400 & 0.00250 & 0.00179 \\ \hline
    20230102\_MTF & 0.00715 & 0.00280 & -0.00050 & 0.00133 & 0.00407 & 0.00009 \\ \hline
    20230102\_LF & 0.01219 & 0.00737 & 0.00235 & 0.00326 & 0.00327 & 0.00170 \\ \hline
    20230102\_MF & 0.00076 & 0.00418 & -0.00330 & 0.00104 & -0.00051 & -0.00050 \\ \hline
    20230113\_UF & 0.00268 & -0.00282 & 0.00026 & -0.00060 & -0.00097 & -0.00076 \\ \hline
    20230113\_MTF & 0.00937 & 0.00262 & 0.00056 & 0.00052 & 0.00165 & -0.00033 \\ \hline
    20230113\_LF & 0.00259 & -0.00239 & 0.00059 & -0.00011 & -0.00155 & -0.00076 \\ \hline
    20230113\_MF & 0.00336 & 0.00251 & 0.00008 & 0.00052 & 0.00044 & -0.00086 \\ \hline
    20230123\_UF & 0.01692 & -0.00231 & 0.00016 & 0.00062 & 0.00431 & -0.00379 \\ \hline
    20230123\_MTF & 0.02145 & -0.00032 & 0.00143 & 0.00262 & 0.00272 & -0.00621 \\ \hline
    20230123\_LF & 0.01565 & -0.00254 & 0.00056 & 0.00009 & 0.00410 & -0.00340 \\ \hline
    20230123\_MF & 0.01882 & 0.00102 & -0.00005 & 0.00134 & 0.00032 & -0.00404 \\ \hline
    \end{tabular}
    }
    \caption{Pearson correlation magnitudes between order-based imbalance OBI and short-horizon realized returns $\tilde{r}$ for each trading day $d$ and filtration scheme $\mathscr{F}$, at horizons $h \in \{1,10,30,50,80,100\}$ seconds (auto-correlation removed). Each entry corresponds to a horizon-specific correlation score $\mathcal{S}^{\rho}_{\text{OBI}\rightarrow \tilde{r}}(\mathscr{F})$ computed after removing auto-correlation effects. Filter types: UF = Unfiltered, LF = Lifetime Filter, MF = Modification Count Filter, MTF = Modification-Time Filter.}
    \label{tab:appendix_corr_on}
\end{table}

\subsection{Discretized Regime Scores across time horizons}
\begin{table}[H]
    \centering
    \resizebox{\columnwidth}{!}{%
    \begin{tabular}{|l|l|l|l|l|l|l|}
    \hline
    \textbf{Date\_Filter} & \textbf{\shortstack{$\mathcal{S}^{\rho}_{\text{OBI}\rightarrow \tilde{r}}(\mathscr{F})$\\(1s)}} & \textbf{\shortstack{$\mathcal{S}^{\mathcal{R}}_{\text{OBI}\rightarrow \tilde{r}}(\mathscr{F})$\\(1s)}} & \textbf{\shortstack{$\mathcal{S}^{\rho}_{\text{OBI}\rightarrow \tilde{r}}(\mathscr{F})$\\(10s)}} & \textbf{\shortstack{$\mathcal{S}^{\mathcal{R}}_{\text{OBI}\rightarrow \tilde{r}}(\mathscr{F})$\\(10s)}} & \textbf{\shortstack{$\mathcal{S}^{\rho}_{\text{OBI}\rightarrow \tilde{r}}(\mathscr{F})$\\(20s)}} & \textbf{\shortstack{$\mathcal{S}^{\mathcal{R}}_{\text{OBI}\rightarrow \tilde{r}}(\mathscr{F})$\\(20s)}} \\ \hline
    20230102\_UF & -14.7367 & 11.5045 & -4.8592 & 9.9368 & -11.9828 & 10.0901 \\ \hline
    20230102\_MTF & -1.4220 & 11.4001 & -0.3291 & 10.6411 & -5.7166 & 9.9782 \\ \hline
    20230102\_LF & -12.2667 & 11.4294 & -3.4627 & 10.2400 & -9.8812 & 10.3134 \\ \hline
    20230102\_MF & 5.8019 & 11.2123 & -0.1979 & 10.2374 & -6.9759 & 10.3247 \\ \hline
    20230113\_UF & -2.7480 & 10.9939 & 3.3443 & 9.9589 & 0.5709 & 9.5452 \\ \hline
    20230113\_MTF & 2.0873 & 11.2029 & -1.4341 & 10.0322 & 1.3986 & 9.6094 \\ \hline
    20230113\_LF & -1.4729 & 11.4112 & 5.2130 & 10.1280 & 0.3083 & 9.7324 \\ \hline
    20230113\_MF & 8.7904 & 10.9055 & -0.5019 & 9.9959 & -4.9251 & 9.3581 \\ \hline
    20230123\_UF & -20.6215 & 12.8027 & 5.5745 & 11.4413 & -7.7442 & 11.0128 \\ \hline
    20230123\_MTF & -26.5211 & 12.7044 & 4.4931 & 11.3200 & -2.4113 & 11.1883 \\ \hline
    20230123\_LF & -15.4226 & 12.7374 & 5.7728 & 11.3674 & -8.5328 & 10.8279 \\ \hline
    20230123\_MF & -11.4672 & 12.1819 & 5.1172 & 10.9107 & -2.9086 & 10.5802 \\ \hline
    \end{tabular}
    }
    \caption{Discrete-regime association scores between order-based imbalance OBI and short-horizon realized returns $\tilde{r}$ across filtration schemes $\mathscr{F}$ and horizons $h \in \{1,10,20\}$ seconds (auto-correlation included). For each trading day $d$, the columns labelled ``CC'' report the regime-based correlation scores $\mathcal{S}^{\rho}_{\text{OBI}\rightarrow \tilde{r}}(\mathscr{F})$, and the columns labelled ``R'' report the regression-based regime scores $\mathcal{S}^{\mathcal{R}}_{\text{OBI}\rightarrow \tilde{r}}(\mathscr{F})$. Filter types: UF = Unfiltered, LF = Lifetime Filter, MF = Modification Count Filter, MTF = Modification-Time Filter, CC = regime-based correlation score, R = regression-based regime score.}
    \label{tab:appendix_regime_off}
\end{table}

\begin{table}[H]
    \centering
    \resizebox{\columnwidth}{!}{%
    \begin{tabular}{|l|l|l|l|l|l|l|}
    \hline
    \textbf{Date\_Filter} & \textbf{\shortstack{$\mathcal{S}^{\rho}_{\text{OBI}\rightarrow \tilde{r}}(\mathscr{F})$\\(1s)}} & \textbf{\shortstack{$\mathcal{S}^{\mathcal{R}}_{\text{OBI}\rightarrow \tilde{r}}(\mathscr{F})$\\(1s)}} & \textbf{\shortstack{$\mathcal{S}^{\rho}_{\text{OBI}\rightarrow \tilde{r}}(\mathscr{F})$\\(10s)}} & \textbf{\shortstack{$\mathcal{S}^{\mathcal{R}}_{\text{OBI}\rightarrow \tilde{r}}(\mathscr{F})$\\(10s)}} & \textbf{\shortstack{$\mathcal{S}^{\rho}_{\text{OBI}\rightarrow \tilde{r}}(\mathscr{F})$\\(20s)}} & \textbf{\shortstack{$\mathcal{S}^{\mathcal{R}}_{\text{OBI}\rightarrow \tilde{r}}(\mathscr{F})$\\(20s)}} \\ \hline
    20230102\_UF & 3.1379 & 7.4766 & -2.6766 & 6.9846 & -10.3337 & 6.9713 \\ \hline
    20230102\_MTF & -3.8815 & 7.0980 & -0.8615 & 6.5966 & -2.5511 & 6.5350 \\ \hline
    20230102\_LF & 4.1918 & 7.3520 & -2.0460 & 6.7774 & -7.2939 & 6.8755 \\ \hline
    20230102\_MF & -5.1195 & 7.4436 & -12.5525 & 7.0353 & -5.7030 & 7.2179 \\ \hline
    20230113\_UF & -22.3374 & 7.6243 & 0.2531 & 7.1001 & -3.4353 & 6.8736 \\ \hline
    20230113\_MTF & -24.9570 & 7.5030 & -5.4746 & 6.8837 & -3.1096 & 6.8727 \\ \hline
    20230113\_LF & -23.1734 & 7.5998 & 0.1417 & 7.0376 & -3.8746 & 6.6639 \\ \hline
    20230113\_MF & -20.0740 & 7.6131 & -11.4961 & 6.8947 & -3.2838 & 6.6064 \\ \hline
    20230123\_UF & -28.2047 & 10.1765 & 3.0024 & 9.3359 & -6.3069 & 8.8993 \\ \hline
    20230123\_MTF & -26.9336 & 8.7339 & -1.7586 & 8.0028 & -3.7381 & 7.7381 \\ \hline
    20230123\_LF & -29.3954 & 9.6022 & 4.2362 & 8.6465 & -7.1072 & 8.1884 \\ \hline
    20230123\_MF & -21.4077 & 9.9060 & 0.8114 & 8.7893 & -3.9746 & 8.5355 \\ \hline
    \end{tabular}
    }
    \caption{Discrete-regime association scores between order-based imbalance OBI and short-horizon realized returns $\tilde{r}$ across filtration schemes $\mathscr{F}$ and horizons $h \in \{1,10,20\}$ seconds (auto-correlation removed). For each trading day $d$, the columns labelled ``CC'' report the regime-based correlation scores $\mathcal{S}^{\rho}_{\text{OBI}\rightarrow \tilde{r}}(\mathscr{F})$, and the columns labelled ``R'' report the regression-based regime scores $\mathcal{S}^{\mathcal{R}}_{\text{OBI}\rightarrow \tilde{r}}(\mathscr{F})$ after auto-correlation removal. Filter types: UF = Unfiltered, LF = Lifetime Filter, MF = Modification Count Filter, MTF = Modification-Time Filter, CC = regime-based correlation score, R = regression-based regime score.}
    \label{tab:appendix_regime_on}
\end{table}

\subsection{Hawkes Kernel Norms Score}
\begin{table}[H]
    \centering
    \resizebox{0.33\columnwidth}{!}{%
    \begin{tabular}{|l|l|}
    \hline
    \textbf{Date\_Filter} & \textbf{$\mathcal{S}^{\phi}_{\text{OBI}\rightarrow \tilde{r}}(\mathscr{F})$} \\ \hline
    20230102\_UF & 8.7967 \\ \hline
    20230102\_MTF & 8.9752 \\ \hline
    20230102\_LF & 8.7149  \\ \hline
    20230102\_MF & 8.7292 \\ \hline
    20230113\_UF & 8.7664  \\ \hline
    20230113\_MTF & 8.8756 \\ \hline
    20230113\_LF & 8.7267  \\ \hline
    20230113\_MF & 8.8249  \\ \hline
    20230123\_UF & 9.2244  \\ \hline
    20230123\_MTF & 9.6726 \\ \hline
    20230123\_LF & 9.3454  \\ \hline
    20230123\_MF & 9.4602  \\ \hline
    \end{tabular}
    }
    \caption{Hawkes excitation norms for order-based imbalance OBI across filtration schemes $\mathscr{F}$ and trading days $d$. Each entry reports the Hawkes-based score $\mathcal{S}^{\phi}_{\text{OBI}\rightarrow \tilde{r}}(\mathscr{F})$, computed as the sum of cross-excitation kernel norms from imbalance to return regimes under a sum-of-exponentials kernel specification. Filter types: UF = Unfiltered, LF = Lifetime Filter, MF = Modification Count Filter, MTF = Modification-Time Filter.}
    \label{tab:appendix_hawkes_obi}
\end{table}

\end{document}